%% file: main.tex
\input{Praeambel}

\begin{document}

\title{Quantum Speed Limits from Symmetries in Quantum Control}

\author{Marco Wiedmann and Daniel Burgarth}

\address{Department Physik, Friedrich-Alexander-Universität Erlangen-Nürnberg, Staudtstraße 7, 91058 Erlangen, Germany}

\begin{abstract}
    In quantum control, quantum speed limits provide fundamental lower bounds on the time that is needed to implement certain unitary transformations. Using Lie algebraic methods, we link these speed limits to symmetries of the control Hamiltonians and provide quantitative bounds that can be calculated without solving the controlled system dynamics. In particular we focus on two scenarios: On one hand, we provide bounds on the time that is needed for a control system to implement a given target unitary \(U\) and on the other hand we bound the time that is needed to implement the dynamics of a target Hamiltonian \(H\) in the worst case. We apply our abstract bounds on physically relevant systems like coupled qubits, spin chains, globally controlled Rydberg atoms and NMR molecules and compare our results to the existing literature. We hope that our bounds can aid experimentalists to identify bottlenecks and design faster quantum control systems.
\end{abstract}
\noindent{\it Keywords\/ Quantum Speed Limit, Quantum Control, Lie Algebras, Symmetries}

\maketitle

\input{Introduction}

\input{Setup}

\input{Result}

\input{Discussion}

\input{Examples}

\input{Conclusion}

\section*{Acknowledgments}
We want to thank Christian Arenz for fruitful discussions on possible physical example systems to consider. This research was supported by the Bavarian Ministry of Economic Affairs, Regional Development and Energy with funds from the Hightech Agenda Bayern and is part of the funding measure Munich Quantum Valley.

\section*{Data availability statement}
All data that support the findings of this study are included within the article.

\section*{Conflicts of Interest Statement}
The authors report there are no competing interests to declare.

\appendix

\input{Appendix}

\printbibliography
\end{document}

%% file: Praeambel.tex
\documentclass[12pt]{iopart}
\usepackage{iopams}
\usepackage[english]{babel}
\usepackage[utf8]{inputenc}
\usepackage[T1]{fontenc}
\usepackage{parskip}
\usepackage{quoting}
\usepackage{microtype}
\usepackage{lmodern}
\usepackage[autostyle]{csquotes}
\usepackage{amsfonts}
\usepackage{amssymb}
\usepackage{dsfont}
\usepackage{acro}
\usepackage{xfrac}
\usepackage{graphicx}
\usepackage{pdfpages}
\usepackage{units}
\usepackage[doi=true, isbn=false, eprint=false, backend=biber, citestyle=nature, sorting=none, giveninits=true, url=false]{biblatex}
\usepackage[nottoc,numbib]{tocbibind}
\usepackage[hypertexnames=false]{hyperref}
\usepackage{cleveref}

\newcommand{\bra}[1]{\langle #1 \rvert}
\newcommand{\ket}[1]{\lvert #1 \rangle}
\newcommand{\braket}[2]{\langle #1 \vert #2 \rangle}

\newcommand{\defeq}{\mathrel{\mathop:}=}

\newcommand{\lVert}{\|}
\newcommand{\rVert}{\|}
\newcommand{\lvert}{|}
\newcommand{\rvert}{|}
\newcommand{\overset}[2]{\mathrel{\mathop{#2}\limits^{#1}}}
\newcommand{\underset}[2]{\mathrel{\mathop{#2}\limits_{#1}}}

\newtheorem{theorem}{Theorem}

\newtheorem{lemma}[theorem]{Lemma}
\newtheorem{corollary}[theorem]{Corollary}

\newtheorem{remark}[theorem]{Remark}

\newenvironment{proof}{
  \par\noindent\textbf{Proof.}\ }
{
\hfill\(\square\)\par}

\newcommand{\ols}[1]{\mskip.5\thinmuskip\overline{\mskip-.5\thinmuskip {#1} \mskip-.5\thinmuskip}\mskip.5\thinmuskip} 
\newcommand{\olsi}[1]{\,\overline{\!{#1}}} 
\makeatletter
\newcommand\closure[1]{
  \tctestifnum{\count@stringtoks{#1}>1} 
  {\ols{#1}} 
  {\olsi{#1}} 
}
\long\def\count@stringtoks#1{\tc@earg\count@toks{\string#1}}
\long\def\count@toks#1{\the\numexpr-1\count@@toks#1.\tc@endcnt}
\long\def\count@@toks#1#2\tc@endcnt{+1\tc@ifempty{#2}{\relax}{\count@@toks#2\tc@endcnt}}
\def\tc@ifempty#1{\tc@testxifx{\expandafter\relax\detokenize{#1}\relax}}
\long\def\tc@earg#1#2{\expandafter#1\expandafter{#2}}
\long\def\tctestifnum#1{\tctestifcon{\ifnum#1\relax}}
\long\def\tctestifcon#1{#1\expandafter\tc@exfirst\else\expandafter\tc@exsecond\fi}
\long\def\tc@testxifx{\tc@earg\tctestifx}
\long\def\tctestifx#1{\tctestifcon{\ifx#1}}
\long\def\tc@exfirst#1#2{#1}
\long\def\tc@exsecond#1#2{#2}
\makeatother

\DeclareAcronym{DLA}{
    short = DLA ,
    long = dynamical Lie algebra
}
\DeclareAcronym{GRAPE}{
    short = GRAPE ,
    long = Gradient Ascent Pulse Engineering
}
\DeclareAcronym{ONB}{
    short = ONB ,
    long = orthonormal basis
}
\DeclareAcronym{NMR}{
    short = NMR ,
    long = nuclear magnetic resonance
}

%% file: Introduction.tex
\section{Introduction}
\label{sec:introduction}
Quantum control and by extension also quantum computations are a race against the clock set by decoherence.
Especially in today's noisy intermediate-scale quantum (NISQ) devices, the quantum system needs to be manipulated and measured before it decoheres too much.
But also in fault tolerant devices, error correction cycles need to be performed fast enough such that errors do not accumulate.
Therefore, there has been a considerable effort in the recent years to engineer faster and faster quantum gates \cite{isenhower_demonstration_2010, jau_entangling_2016, levine_parallel_2019, evered_high-fidelity_2023, paraoanu_microwave-induced_2006, corcoles_process_2013, sheldon_procedure_2016, howard_implementing_2023}.
However, this strategy is doomed to hit a wall imposed by the fundamental laws of quantum mechanics: the quantum speed limit.

In their original formulation, quantum speed limits bound the time that a quantum system governed by a time-independent Hamiltonian needs to evolve a quantum state into an orthogonal state \cite{mandelstam_uncertainty_1991, margolus_maximum_1998}.
The concept can be adapted to quantum control theory. There, it usually denotes the minimal duration of the control pulse needed to reach a certain target.
The existence of such a minimal duration has been studied both analytically and numerically in the past \cite{vidal_interaction_2002, caneva_optimal_2009, arenz_roles_2017, lee_dependence_2018, burgarth_quantum_2022, damme_motion-insensitive_2024}.
However, numerical methods require solving the time-dependent dynamics of the control system, which quickly becomes intractable for larger systems.

It has recently been established that quantum speed limits can arise from the distance to uncontrollability \cite{burgarth_quantum_2022}. Intuitively speaking, if a controllable system is close to an uncontrollable one, we expect the dynamics of the two to stay close to each other at small timescales, meaning the controllability of the former system needs time to express itself.
We generalize this concept to a \textit{distance to unreachability}, where we consider perturbations that make a \emph{specific} target unattainable instead. 

In order to characterize the reachability of the target operation in different control systems, we employ methods based on the quadratic symmetries of the \ac{DLA} \cite{zeier_symmetry_2011, zeier_squares_2015, zimboras_symmetry_2015}, therefore establishing a link between the quantum speed limit and said symmetries.
The bounds we obtain crucially depend on the size of the commutators between a symmetry and both the target operation and the control system Hamiltonian, quantifying the intuitive idea that the \enquote{rate of symmetry breaking} is a bottleneck in quantum control.

The rest of this paper is structured as follows: After \cref{sec:setup} gives a short introduction to relevant concepts from quantum control, \cref{sec:main_result} states the main results of this paper with short proofs. Details are deferred to the appendix.
\Cref{sec:discussion} discusses important properties of the obtained bounds. Concrete examples are given and discussed in \cref{sec:speed_limit_application}. Finally, we conclude in \cref{sec:conclusion}.

%% file: Setup.tex
\section{Setup}
\label{sec:setup}

Consider a bilinear quantum control system on a \(d\)-dimensional (\(d < \infty\)) Hilbert space \(\mathcal{H}\) of the form
\begin{equation}
    H(t) = H_d + \sum_{j=1}^n f_j(t) H_j,
\end{equation}
with an always-on drift term \(H_d\), control Hamiltonians \(H_j\) and corresponding control functions \(f_j(t)\).
For now let us assume that the controls can in principle be engineered to be arbitrarily strong, i.e. we impose no bounds on the control functions. This means that the quantum speed limit is set by the drift --- a realistic scenario in many applications where the bottleneck is the weak couplings between qubits.
Given any choice for the \(f_j(t)\), the Schrödinger equation is formally solved by the unitary operator
\begin{equation}
    U(t) = \mathcal{T}\exp\left(-i \int_0^t \mathrm{d}s H(s)\right),
\end{equation}
where \(\mathcal{T}\) denotes the time-ordering operator.

The reachable set \(\mathcal{R}\) of the control system is the set of all unitaries \(V\) for which there exists a choice of control functions such that \(U(t) = V\) for some time \(t\).
It can be characterized by the \ac{DLA} \(\mathfrak{g}\) of the control system, which is the Lie closure of the drift and control Hamiltonians
\begin{equation}
    \mathfrak{g} = \langle iH_d, iH_1, ..., iH_n \rangle_\text{Lie}.
\end{equation}
The reachable set is dense in the exponential of the \ac{DLA} \cite{jurdjevic_control_1972}
\begin{equation}
    \label{eq:reachable_set}
    \closure{\mathcal{R}} = \exp(\mathfrak{g}),
\end{equation}
where the closure is taken with respect to the subspace topology of \(\mathrm{U}(d)\).
Since all Lie subalgebras of \(\mathfrak{su}(d)\) are reductive and the abelian components can be treated separately, we assume that the \ac{DLA} is semisimple.
As any semisimple subalgebra of a compact algebra is itself compact, it follows that also the \ac{DLA} is compact.
In that case the closure in \eref{eq:reachable_set} is no longer necessary. Further, this guarantees the existence of a finite, uniform bound on the control time, i.e. there exists a \(T_* < \infty\) such that any unitary in \(\mathcal{R}\) can be exactly reached in \(t < T_*\) units of time.

It has previously been established in \cite{burgarth_quantum_2022} that for universal systems, i.e. when \(\mathcal{R} = \mathrm{SU}(d)\), \(T_*\) can be bounded from below by 
\begin{equation}
    \label{eq:distance_to_uncontrollability}
    T_* \geq \frac{1}{4 \lVert \Delta H \rVert_\infty},
\end{equation}
where \(\Delta H\) is a perturbation to the drift that renders the control system non-universal and \(\lVert \cdot \rVert_\infty\) is the operator norm, i.e. the largest modulus of the eigenvalues.
However, in most practical situations, one is not necessarily interested in \(T_*\) itself.
It could for example be the case that the unitaries in \(\mathcal{R}\) that saturate the uniform control time are not practically relevant. At the same time, there are always unitaries which can be reached very quickly (a trivial example being the identity). This calls for target-dependent bounds.
Hence, we generalize the methods from \cite{burgarth_quantum_2022} for two possible applications.
\begin{enumerate}
    \item One wants to implement a specific unitary operation \(U\) on a control system and wants to find out how much time is needed at least to do so.
    \item One wants to use a control system to simulate the time evolution of a particular Hamiltonian \(H_s\) and wants to find out how much time is needed at least to reach any unitary in its orbit.
\end{enumerate}

Note that if the target can be reached through the controls alone, i.e. if
\begin{equation}
    U \in \exp(\mathfrak{h}) \text{ or } H_s \in \mathfrak{h} \text{ for } \mathfrak{h} = \langle iH_1, ..., iH_n \rangle_\text{Lie},
\end{equation}
then it can be reached arbitrarily fast by appropriately large controls.
Therefore we want to restrict to control systems where \(\mathfrak{h} \subset \mathfrak{g}\) is a strict Lie subalgebra and to targets which are not reachable with \(\mathfrak{h}\) alone.
Proper subalgebras of a compact semisimple Lie algebras admit a larger set of quadratic symmetries \cite{zeier_squares_2015}.
These are matrices \(S\) on \(\mathcal{H}^{\otimes 2}\), which commute with \(X \otimes \mathds{1} + \mathds{1} \otimes X\) for any element \(X\) of the Lie algebra.
Note that any symmetry \(S\) in the commutant of \(\mathfrak{h}\) corresponds to a quadratic symmetry \(S \otimes \mathds{1} + \mathds{1} \otimes S\).
In the following we will make use of the fact that \(\mathfrak{h}\) admits quadratic symmetries that are not shared by \(\mathfrak{g}\) to derive the quantum speed limits.

%% file: Result.tex
\section{Main results}
\label{sec:main_result}
First, consider the case that one wants to implement a specific target unitary \(U\) with the control system.
Let \(\Delta H\) be a perturbation such that the perturbed control system \(H(t) = H_d + \Delta H + \sum_{j=1}^n f_j(t) H_j\) can no longer reach reach \(U\), i.e. \(U \notin \mathcal{R}'\), where \(\mathcal{R}'\) is the reachable set of the perturbed system.
From Duhamels principle (refer to \ref{app:duhamel}) we obtain that the minimal time needed to implement \(U\) is bounded by
\begin{equation}
    \label{eq:abstract_speed_limit}
    T \geq \frac{1}{\lVert \Delta H \rVert_\infty} \inf_{V \in \mathcal{R}'} \lVert U - V \rVert_\infty.
\end{equation}

Computing the quantities on the right hand side of \eref{eq:abstract_speed_limit} is a hard problem in general.
Therefore we instead further bound the right hand side of \eref{eq:abstract_speed_limit} in terms of quantities that are easier to calculate.

\begin{lemma}
    \label{lem:unitary_delta}
    Let S be a quadratic symmetry of \(\mathfrak{h}\) that is not shared by \(\mathfrak{g}\). Further, let \(\Delta H\) be a perturbation, such that \(\mathfrak{g}' = \langle H_d + \Delta H, H_1, ..., H_n \rangle_\text{Lie}\) respects \(S\). Then
    \begin{equation}
        \inf_{V \in \mathcal{R}'} \lVert U - V \rVert_\infty \geq \frac{\lVert [U^{\otimes 2}, S] \rVert_F}{4 \lVert S \rVert_F},
    \end{equation}
    where \(\lVert \cdot \rVert_F\) denotes the Frobenius norm.
\end{lemma}
\begin{proof}
    For any \(V \in \mathcal{R}'\) we have that
    \begin{equation}
        \eqalign{
            \lVert U^{\otimes 2} - V^{\otimes 2} \rVert_\infty &\leq \lVert (U \otimes U - U \otimes V \rVert_\infty
            + \lVert U \otimes V -  V \otimes V \rVert_\infty \\
            &= 2 \lVert U - V \rVert_\infty.
        }
    \end{equation}
    Similarly, through the triangle inequality we get that
    \begin{equation}
        \eqalign{
            \fl \lVert U^{\otimes 2} - V^{\otimes 2} \rVert_\infty \geq \frac{1}{2} \lVert (U^{\otimes 2} - V^{\otimes 2})\otimes \mathds{1} - \mathds{1} \otimes (U^{\otimes 2} - V^{\otimes 2})^T \rVert_\infty \\
            = \frac{1}{2} \lVert \mathrm{ad}_{U^{\otimes 2}} - \mathrm{ad}_{V^{\otimes 2}} \rVert_\infty \geq \frac{1}{2 \lVert S \rVert_F} \lVert \mathrm{ad}_{U^{\otimes 2}}(S) - \mathrm{ad}_{V^{\otimes 2}}(S) \rVert_F,
        }
    \end{equation}
    where the last inequality is due to the definition of the operator norm. Here, we used \(\mathrm{ad}_X(Y) \defeq [X, Y] = (X \otimes \mathds{1} - \mathds{1}\otimes X^T)\text{vec}(Y)\) by Roth's lemma, where \(\text{vec}\) denotes the row-vectorization, i.e. the vector obtained by stacking the rows of a matrix. Note that the Frobenius norm is the same as the Euclidean norm of the vectorization \(\lVert X \rVert_F = \lVert \mathrm{vec}(X) \rVert_2\).
    Since \(V \in \mathcal{R}'\) and \(S\) is a quadratic symmetry of \(\mathfrak{g}'\), it follows that \(\mathrm{ad}_{V^{\otimes 2}}(S) = 0\).
    The statement immediately follows.
\end{proof}

\begin{corollary}
    Let S be a symmetry of \(\mathfrak{h}\) that is not shared by \(\mathfrak{g}\). Further, let \(\Delta H\) be a perturbation, such that \(\mathfrak{g}' = \langle H_d + \Delta H, H_1, ..., H_n \rangle_\text{Lie}\) respects \(S\). Then
        \begin{equation}
        \inf_{V \in \mathcal{R}'} \lVert U - V \rVert_\infty \geq \frac{\lVert [U, S] \rVert_F}{2 \lVert S \rVert_F}.
    \end{equation}
\end{corollary}

Note that the (quadratic) symmetries of a control system can be found by solving a linear system of equations \cite{burgarth_quantum_2022}.

As we see from \cref{lem:unitary_delta}, when \([U, S] \neq 0\), then by restoring the symmetry \(S\) to the control system with the perturbation \(\Delta H\) we render \(U\) unreachable to the control system.
How closely the perturbed system can still approximate \(U\) only depends on how much \(S\) is broken by \(U\), i.e. the norm of the commutator \([U, S]\).

Next, consider that one wants to study the dynamics of some Hamiltonian \(H_s\) at various times.
From the compactness of the \ac{DLA} we know that any unitary in the orbit of \(H_s\), \(U(t) = \exp(-iH_st)\) can be reached in a finite amount of time.
We ask the question: how much time \(T\) is needed in the worst case?
Similarly as before, from Duhamels principle we get that
\begin{equation}
    \label{eq:abstract_simulation_speed_limit}
    T \geq \frac{1}{\lVert \Delta H \rVert_\infty} \sup_{t \geq 0} \inf_{V \in \mathcal{R}'} \lVert e^{-iH_st} - V \rVert_\infty,
\end{equation}
where \(\Delta H\) now is chosen such that \(H_s \notin \mathfrak{g}'\).
The following lemma establishes how to bound the supremum on the right hand side.

\begin{lemma}
    \label{lem:hamiltonian_lower_bound}
    Let S be a quadratic symmetry of \(\mathfrak{h}\) that is not shared by \(\mathfrak{g}\).
    Further, let \(\Delta H\) be a perturbation such that \(\mathfrak{g}' = \langle H_d +
     \Delta H, H_1, ..., H_n \rangle_\text{Lie}\) respects \(S\). Then
     \begin{equation}
         \sup_{t \geq 0} \inf_{V \in \mathcal{R}'} \lVert e^{-iH_st} - V \rVert_\infty \geq \frac{\lVert (\mathds{1} - P_{\mathrm{ker} \, \mathrm{ad}_{H_s \otimes \mathds{1} + \mathds{1} \otimes H_s}}) S \rVert_F}{2\sqrt{2} \lVert S \rVert_F},
     \end{equation}
     where \(P_{\mathrm{ker} \, \mathrm{ad}_{H_s \otimes \mathds{1} + \mathds{1} \otimes H_s}}\) is the orthogonal projector onto the kernel of \(\mathrm{ad}_{H_s \otimes \mathds{1} + \mathds{1} \otimes H_s}\).
\end{lemma}

\begin{proof}
    By applying the supremum on both sides of \cref{lem:unitary_delta} and using the unitary invariance of the Frobenius norm, we get that
    \begin{equation}
        \sup_{t \geq 0} \inf_{V \in \mathcal{R}'} \lVert e^{-iH_st} - V \rVert_\infty \geq \sup_{t \geq 0} \frac{\lVert \mathrm{Ad}_{U^{\otimes 2}}(S) - S \rVert_F}{4  \lVert S \rVert_F},
    \end{equation}
    where \(\mathrm{Ad}_X(Y) = XYX^\dagger\) is the adjoint action.
    Using that \(\mathrm{Ad}_{e^{-iXt}} = e^{-it\mathrm{ad}_{X}}\) for any Hermitian \(X\) and \(\exp(-iH_st)^{\otimes 2} = \exp(-i (H_s \otimes \mathds{1} + \mathds{1} \otimes H_s)t)\), it follows that 
    \begin{equation}
        \sup_{t \geq 0} \frac{\lVert \mathrm{Ad}_{U^{\otimes 2}}(S) - S \rVert_F}{4  \lVert S \rVert_F} 
        \geq  \sup_{t \geq 0} \frac{\lVert \left(\exp(-i \mathrm{ad}_{H_s \otimes \mathds{1} + \mathds{1} \otimes H_s}t) - \mathds{1}\right) S \rVert_F}{4 \lVert S \rVert_F}.
    \end{equation}
    According to \cref{lem:adjoint_hermitian}, \(\mathrm{ad}_{H_s \otimes \mathds{1} + \mathds{1} \otimes H_s}\) is Hermitian, therefore we can apply \cref{lem:lower_bound_evolution_from_identity} from \ref{app:lemmas} and the result immediately follows.
\end{proof}

\begin{corollary}
    Let S be a symmetry of \(\mathfrak{h}\) that is not shared by \(\mathfrak{g}\).
    Further, let \(\Delta H\) be a perturbation such that \(\mathfrak{g}' = \langle H_d +
     \Delta H, H_1, ..., H_n \rangle_\text{Lie}\) respects \(S\). Then
     \begin{equation}
         \sup_{t \geq 0} \inf_{V \in \mathcal{R}'} \lVert e^{-iH_st} - V \rVert_\infty \geq \frac{\lVert (\mathds{1} - P_{\mathrm{ker} \, \mathrm{ad}_{H_s}}) S \rVert_F}{\sqrt{2} \lVert S \rVert_F}.
     \end{equation}
\end{corollary}

\begin{remark}
    In practice, computing \(P_{\mathrm{ker}\, \mathrm{ad}_{H_s}}\) involves diagonalizing the simulation Hamiltonian \(H_s\). However, this would render the actual simulation of \(H_s\) unnecessary, since the classical computer could just solve the time evolution explicitly at that point.
    Therefore, we propose two techniques to further bound the norm of the projection.
    \begin{enumerate}
        \item Using \cref{lem:lower_bound_projection_to_kernel} from \ref{app:lemmas}, we bound the projection in terms of the commutator between \(H_s\) and \(S\)
        \begin{equation}
            \lVert (\mathds{1} - P_{\mathrm{ker} \, \mathrm{ad}_{H_s}}) S \rVert_F \geq \frac{\lVert [H_s, S] \rVert_F}{2 \lVert H_s \rVert_\infty},
        \end{equation}
        and similarly for quadratic symmetries.
        While this bound is easier to calculate in many cases, it can be considerably loose.
        \item If the smallest and largest gap between eigenvalues of \(H_s\) are known, then one can numerically approximate the projection onto the kernel as a Chebyshev polynomial of \(\mathrm{ad}_{H_s}\), see \ref{app:chebyshev} for more details.
    \end{enumerate}
\end{remark}

Now it only remains to find suitable bounds on the size of the perturbation \(\Delta H\) that restores a quadratic symmetry \(S\) to the original control system.
Denote by \(\iota\) the mapping to the squared Hilbert space
\begin{eqnarray}
        \iota: \mathfrak{su}(d) & \rightarrow \mathfrak{su}(d^2) \\
        iH & \mapsto (iH) \otimes \mathds{1} + \mathds{1} \otimes (iH),
\end{eqnarray}
and define \(A \defeq \mathrm{ad}_{iS} \circ \iota\).
Then, \(\Delta H\) needs to solve the linear equation
\begin{equation}
    \label{eq:commutator_equation}
    A(i\Delta H) = - A(iH_d).
\end{equation}
Note that solutions always exist, as \(\Delta H = - H_d\) solves \eref{eq:commutator_equation}.
For small systems, \(A\) and its Moore-Penrose pseudoinverse \(A^+\) with \(AA^+A=A\) can be computed numerically in some basis.
Then, \(i\Delta H \defeq -A^+A(iH_d)\) solves \eref{eq:commutator_equation} and we get the inequality
\begin{equation}
    \lVert \Delta H \rVert_\infty \leq \lVert \Delta H \rVert_F \leq \frac{1}{\sigma_\mathrm{min}} \lVert [S, \iota(H_d)] \rVert_F,
\end{equation}
where \(\sigma_\mathrm{min}\) denotes the smallest non-zero singular value of \(A\).
As far as we are aware, it is not possible to estimate \(\sigma_\mathrm{min}\) without explicitly constructing \(A\).
In the case of a linear symmetry there is no need to construct \(A\), as the following Lemma shows.
\begin{lemma}
     Let \(S\) be a symmetry of \(\mathfrak{h}\) that is not shared by \(\mathfrak{g}\). Then there exists a perturbation \(\Delta H\) such that \(\mathfrak{g}' = \langle H_d +
     \Delta H, H_1, ..., H_n \rangle_\text{Lie}\) respects \(S\) and that is bounded by
     \begin{equation}
         \lVert \Delta H \rVert_\infty \leq \frac{\lVert[S, H_d]\rVert_F}{\sigma_\text{min}},
     \end{equation}
     where \(\sigma_\text{min} = \min\{\lvert \lambda_i - \lambda_j \rvert | \lambda_i, \lambda_j \in \sigma(S), \lambda_i \neq \lambda_j\}\).
\end{lemma}
\begin{proof}
    \(\Delta H\) restores the quadratic symmetry \(S\) to \(\mathfrak{g}'\) if and only if it solves the equation
    \begin{equation}
        \label{eq:ad_equation}
        \mathrm{ad}_{S} (\Delta H) = -\mathrm{ad}_{S}(H_d).
    \end{equation}
    Denote by \(\mathrm{ad}_{S}^+\) the Moore-Penrose pseudoinverse of \(\mathrm{ad}_S\).
    By its properties, \(\Delta H \defeq - \mathrm{ad}_{S}^+ \mathrm{ad}_S(H_d)\) is the unique solution to \eref{eq:ad_equation} with minimal Frobenius norm.
    Since \((\Delta H)^\dagger\) also solves \eref{eq:ad_equation} and the Frobenius norm is invariant under Hermitian conjugation, we conclude that \(\Delta H\) is indeed Hermitian.

    The spectrum of \(\mathrm{ad}_S\) is well-known to be \(\sigma(\mathrm{ad}_S) = \{\lambda_i - \lambda_j | \lambda_i, \lambda_j \in \sigma(S)\}\).
    Since the Moore-Penrose pseudoinverse is bounded by the reciprocal of the smallest nonzero eigenvalue of \(\mathrm{ad}_S\), the statement follows.
\end{proof}

\begin{remark}
    \(S\) can always be chosen to be a projection, in which case \(\sigma_\text{min} = 1\).
\end{remark}

The efforts of this section are summarized by the following theorems, which constitute the main results of this paper.
\begin{theorem}
\label{thm:speed_limit_unitary}
Let \(U \in \mathcal{R}\) be a reachable unitary.
\begin{itemize}
    \item [(a)] Let \(S\) be a quadratic symmetry of \(\mathfrak{h}\), that is not shared by \(\mathfrak{g}\) and \(\sigma_\mathrm{min} = \min\{\lvert \lambda_i - \lambda_j \rvert | \lambda_i, \lambda_j \in \sigma(S), \lambda_i \neq \lambda_j\}\). Further, let \(\Delta H\) be a perturbation such that \(\mathfrak{g}' = \langle H_d + \Delta H, H_1, ..., H_n \rangle_\text{Lie}\) respects \(S\).
    Then, the minimal time \(T\) needed to implement \(U\) is bounded from below by
    \begin{equation}
        T \geq \frac{\lVert [U^{\otimes 2}, S] \rVert_F}{4  \lVert S \rVert_F \lVert \Delta H \rVert_\infty}.
    \end{equation}
    \item[(b)] Let \(S\) be a symmetry of \(\mathfrak{h}\), that is not shared by \(\mathfrak{g}\) and \(\sigma_\mathrm{min} = \min\{\lvert \lambda_i - \lambda_j \rvert | \lambda_i, \lambda_j \in \sigma(S), \lambda_i \neq \lambda_j\}\). Further, let \(\Delta H\) be a perturbation such that \(\mathfrak{g}' = \langle H_d + \Delta H, H_1, ..., H_n \rangle_\text{Lie}\) respects \(S\).
    Then, the minimal time \(T\) needed to implement \(U\) is bounded from below by
    \begin{equation}
        T \geq \frac{\lVert [U, S] \rVert_F}{2 \lVert S \rVert_F \lVert \Delta H \rVert_\infty} \geq \frac{\lVert [U, S] \rVert_F \sigma_\mathrm{min}}{2 \lVert S \rVert_F \lVert [S, H_d] \rVert_F}.
    \end{equation}
\end{itemize}
\end{theorem}

\begin{theorem}
    \label{thm:speed_limit_hamiltonian}
    Let \(iH_s \in \mathfrak{g}\) be a simulable Hamiltonian.
    \begin{itemize}
        \item [(a)] Let \(S\) be a quadratic symmetry of \(\mathfrak{h}\), that is not shared by \(\mathfrak{g}\) and \(\sigma_\mathrm{min} = \min\{\lvert \lambda_i - \lambda_j \rvert | \lambda_i, \lambda_j \in \sigma(S), \lambda_i \neq \lambda_j\}\). Further, let \(\Delta H\) be a perturbation such that \(\mathfrak{g}' = \langle H_d + \Delta H, H_1, ..., H_n \rangle_\text{Lie}\) respects \(S\).
        Then, there exists a unitary \(U\) in the orbit of \(H_s\) such that the minimal time \(T\) needed to implement \(U\) is bounded from below by
        \begin{equation}
            T \geq \frac{\lVert (\mathds{1} - P_{\mathrm{ker} \, \mathrm{ad}_{H_s \otimes \mathds{1} + \mathds{1} \otimes H_s}}) S \rVert_F}{2\sqrt{2} \lVert S \rVert_F \lVert \Delta H \rVert_\infty} \geq \frac{\lVert [H_s \otimes \mathds{1} + \mathds{1} \otimes H_s, S] \rVert_F}{8 \sqrt{2} \lVert S \rVert_F \lVert H_s \rVert_\infty \lVert \Delta H \rVert_\infty}.
        \end{equation}
        \item[(b)] Let \(S\) be a symmetry of \(\mathfrak{h}\), that is not shared by \(\mathfrak{g}\) and \(\sigma_\mathrm{min} = \min\{\lvert \lambda_i - \lambda_j \rvert | \lambda_i, \lambda_j \in \sigma(S), \lambda_i \neq \lambda_j\}\). Further, let \(\Delta H\) be a perturbation such that \(\mathfrak{g}' = \langle H_d + \Delta H, H_1, ..., H_n \rangle_\text{Lie}\) respects \(S\).
        Then, there exists a unitary \(U\) in the orbit of \(H_s\) such that the minimal time \(T\) needed to implement \(U\) is bounded from below by
        \begin{equation}
            \eqalign{
                T \geq \frac{\lVert (\mathds{1} - P_{\mathrm{ker} \, \mathrm{ad}_{H_s}}) S \rVert_F}{\sqrt{2} \lVert S \rVert_F \lVert \Delta H \rVert_\infty}
                & \geq \frac{\lVert [H_s, S] \rVert_F}{2 \sqrt{2} \lVert S \rVert_F \lVert H_s \rVert_\infty \lVert \Delta H \rVert_\infty} \nonumber \\
                & \geq \frac{\lVert [H_s, S] \rVert_F \sigma_\mathrm{min}}{2\sqrt{2} \lVert S \rVert_F \lVert H_s \rVert_\infty \lVert [S, H_d] \rVert_F}.
            }
        \end{equation}
\end{itemize}
\end{theorem}

%% file: Discussion.tex
\section{Discussion}
\label{sec:discussion}
Unlike in \eref{eq:distance_to_uncontrollability}, the right hand sides of \eref{eq:abstract_speed_limit} and \eref{eq:abstract_simulation_speed_limit} explicitly depend on the choices of \(U\) or \(H_s\) and \(\mathcal{R}'\).
In \cite{burgarth_quantum_2022} it was proven that
\begin{equation}
    \sup_{U \in \mathrm{SU}(d)} \inf_{V \in \mathcal{R}` \subsetneq \mathrm{SU}(d)} \lVert U - V \rVert_\infty \geq \frac{1}{4},
\end{equation}
for any strict Lie subgroup \(\mathcal{R}' \subsetneq \mathrm{SU}(d)\).
We discuss in the following why such a uniform bound cannot be derived when restricting oneself to a particular target.

It is obvious that for any strict Lie subgroup \(\mathcal{R}'\) it is possible to find some target unitary that lies \(\epsilon\)-close to, but outside of \(\mathcal{R}'\) for any choice of \(\epsilon > 0\).
The more interesting question is: can we find for every strict Lie subgroup \(\mathcal{R}' \subsetneq \mathrm{SU}(d)\) a Hamiltonian \(H_s\), such that the whole orbit of \(H_s\) stays \(\epsilon\)-close to \(\mathcal{R}'\) but is not contained in \(\mathcal{R}'\)?
The following Theorem gives an affirmative answer to this question.

\begin{theorem}
\label{thm:no_model_independence}
For all non-trivial proper Lie subgroups \(\mathcal{R}' \subsetneq \mathrm{SU}(d)\), and for all \(\epsilon > 0\) there exists a target Hamiltonian \(H_s\) such that the target cannot be fully simulated by \(\mathcal{R}'\), i.e., \(\{e^{-iH_st} | t > 0\} \nsubseteq \mathcal{R}'\), but any unitary time evolution \(U\) in the orbit of \(H_s\) can be approximated \(\epsilon\)-close by some \(V \in \mathcal{R}'\).
\end{theorem}
\begin{proof}
	Let \(\epsilon > 0\), and \(\mathfrak{g}'\) be the Lie algebra of \(\mathcal{R}'\).
	Since \(\mathfrak{su}(d)\) is simple, there exist Hamiltonians \(i H \in \mathfrak{su}(d)\) and \(i X \in \mathfrak{g}'\) such that \(\left[i H, i X\right] \notin \mathfrak{g}'\).
	Otherwise, \(\mathfrak{g}'\) would be an ideal of \(\mathfrak{su}(d)\) and \(\mathfrak{g}'\) would need to be trivial.
	Consider now the Hamiltonian obtained by slightly rotating \(X\) with \(H\)
	\begin{equation}
		H_s^{(\theta)} \defeq e^{-iH\theta}Xe^{iH\theta} \in \mathfrak{su}(d).
	\end{equation}
	The proof will conclude by showing that for proper choices of \(\theta\), \(H_s^{(\theta)}\) does not belong to \(\mathcal{R}'\) and can indeed be simulated \(\epsilon\)-close at all times by \(X\).
	The latter follows from \cref{lem:rotated_unitary_bound} from \ref{app:lemmas} by choosing
    \begin{equation}
        \theta < \frac{\sqrt{1 + \epsilon} - 1}{\rVert H \lVert_\infty}.
    \end{equation}
	Assuming that \(i H_s^{(\theta)} \in \mathfrak{g}'\) for all \(\theta \in [0, \sfrac{(\sqrt{1 + \epsilon} - 1)}{\rVert H \lVert_\infty}]\) would mean that \(\frac{\mathrm{d}}{\mathrm{d}\theta} i H_s^{(\theta)} \mid_{\theta = 0} \in \mathfrak{g}'\).
	However, the derivative evaluates to
	\begin{equation}
		\frac{\mathrm{d}}{\mathrm{d}\theta} i H_s^{(\theta)} \mid_{\theta = 0} = [H, X],
	\end{equation}
	which leads to a contradiction.
\end{proof}

This shows that there cannot be any meaningful universal bounds to the right hand sides of \eref{eq:abstract_speed_limit} and \eref{eq:abstract_simulation_speed_limit}.
The reason for this lies fundamentally within the task that is being considered.
Whenever the target operation can be chosen arbitrarily, one can rely on the fact that there will always be a \(U \in \text{SU}(d)\) that moves states far away from any trajectories reachable in \(\mathcal{R}'\) \cite{burgarth_quantum_2022}.
However, by restricting the targets to a specific set this guarantee is lost.

It is still an open question however, whether bounds exist that do not explicitly depend on the choice of \(\mathcal{R}'\), or equivalently on the choice of symmetry \(S\).
When evaluating the speed limits in \cref{thm:speed_limit_unitary} and \cref{thm:speed_limit_hamiltonian}, one still has some freedom of choice regarding the symmetry \(S\).
While simply rescaling \(S \mapsto \lambda S\), \(\lambda \in \mathds{R}\) does not change the bounds, adding or subtracting multiples of the identity \(S \mapsto S + \lambda \mathds{1}\), \(\lambda \in \mathds{R}\) does.
Also, there might even be multiple linearly independent choices for \(S\), in which case every linear combination would also be a suitable candidate.
Therefore, to make the bounds as tight as possible, one should always try to maximize the bounds given in \cref{thm:speed_limit_unitary} and \cref{thm:speed_limit_hamiltonian} with respect to the choice of \(S\).

%% file: Examples.tex
\section{Application to physical example systems}
\label{sec:speed_limit_application}
We present here a number of examples that illustrate how \cref{thm:speed_limit_unitary} and \cref{thm:speed_limit_hamiltonian} can be applied to physical control systems.

\subsection{CNOT gate on coupled qubits}
\label{sec:speed_limit_cnot}

As a pedagogical example, consider a pair of qubits \(\mathcal{H} = \mathds{C}^2 \otimes \mathds{C}^2\) which are coupled through a Pauli interaction with coupling strength \(g\), i.e. \(H_d = g Z_1 Z_2\).
Assume further access to local control over both of the two qubits, i.e., \(H_1 = X_1 \text{, } H_2 = Z_1 \text{, } H_3 = X_2 \text{, } H_4 = Z_2\).
The goal is to implement a CNOT gate on this system.

As the controls are local, they cannot generate any entanglement. Nonetheless, there is no simple symmetry. We can, however, find a quadratic symmetry
\begin{equation}
    S = \mathds{1}_{16} - M_{(1, 3)} - M_{(2,4)} + M_{(1, 3)(2,4)}
\end{equation}
that relates to the concurrence of a pure two-qubit state \cite{zimboras_symmetry_2015}. Here, $M_{(i,j)}$ is the matrix that permutes the tensor components \(i\) and \(j\). Obviously, the perturbation \(\Delta H = - g Z_1 Z_2\) restores this symmetry to the control system.

By applying \cref{thm:speed_limit_unitary}(a), we get that the minimal time needed to implement a CNOT gate is lower bounded by
\begin{equation}
    T \geq \frac{\sqrt{2}}{4g}.
\end{equation}
For simple systems like this one, the quantum speed limit can be determined exactly. In this case it is known to be \(T_\text{CNOT} = \frac{\pi}{4g}\) \cite{vidal_interaction_2002}, which differs from the above bound only by the multiplicative constant \(\frac{\pi}{\sqrt{2}} \approx 2.22\).
Most importantly, it features the same asymptotic \(\frac{1}{g}\) behavior, showing that the speed of two-qubit gates is fundamentally limited by their interaction strength, no matter which local controls are applied to them.

While the presented speed limit is not tight, the real strength of the methods developed in this paper lies in their versatility.
The following examples demonstrate this strength by calculating bounds on the speed limit on quantum systems where the exact speed limit is still unknown.

\subsection{SWAP gate on hopping chain}
\label{sec:speed_limit_swap}

For the second example consider a single spinless particle on a chain with \(N \in \mathds{N}\) sites and nearest-neighbor coupling between these sites.
The Hilbert space in this example is \(\mathcal{H} = \mathds{C}^N\), to describe the probability amplitude of a particle being located at each site.
The system evolves under a nearest-neighbor coupling described by the Hamiltonian
\begin{equation}
	H_d = J \sum_{j = 1}^{N-1} \left(\ket{j}\bra{j + 1} + \ket{j+1}\bra{j}\right).
\end{equation}
Further, the potential on one end of the chain serves as the control of the system, i.e.,
\begin{equation}
	H_1 = \ket{1}\bra{1}.
\end{equation}

It is known that this control is already enough to render the system fully controllable \cite{wang_symmetry_2012, burgarth_zero_2013}.
Therefore it can implement a swap operation between the two ends of the chain
\begin{equation}
	\text{SWAP} = \mathds{1} - \ket{1}\bra{1} - \ket{N}\bra{N} + \ket{1}\bra{N} + \ket{N}\bra{1}.
\end{equation}
Physically such a SWAP can be useful in the context of quantum state transfer and remote gates. The speed limit to implement it has been studied analytically and numerically in \cite{lee_dependence_2018, blankref}.
While analytically a speed limit of \(T_{\text{SWAP}} \geq \frac{1}{J}\) was found, numerical optimization of the bound showed that the time needed to implement the SWAP operation should grow with the system size.
This is in agreement with explicit numerical calculations of the control pulses through \ac{GRAPE}.

Furthermore it is known that the time needed to implement the most difficult unitaries in this control system scales at least quadratically with the length of the chain \cite{burgarth_quantum_2022}.
However, it is not known what exactly the most difficult unitaries are in this case.
The SWAP operation does provide a natural candidate, as the control has to first propagate through the whole chain before it can affect the other end.

We make use of the fact that the drift Hamiltonian has an energy gap, that closes quadratically with \(N\) to construct a symmetry, that is restored to the drift by closing that gap. For details on the construction, refer to \ref{app:swap}. By applying \cref{thm:speed_limit_unitary}(b), we arrive at the speed limit

\begin{eqnarray}
	T \geq \frac{2 N^2}{3\pi^2 J \sqrt{3 + 2\cos\left(\frac{2 \pi}{N + 1}\right)}} \sqrt{\frac{2}{N+1}} \left \lvert \sin\left(\frac{2 \pi N}{N+1}\right) \right\rvert.
\end{eqnarray}

Asymptotically, the speed limit scales as \(\mathcal{O}(\sqrt{N})\).
It is bounded from below by
\begin{equation}
    T \geq \frac{\sqrt{N}}{19J}.
\end{equation}
This is an asymptotic improvement by a factor of \(\sqrt{N}\) compared to the bound obtained in \cite{lee_dependence_2018}. Remarkably, this bound on the SWAP time explicitly scales with the system size, without normalizing the Hamiltonian.
However, it is not yet close to the \(N^2\) behavior, which is to be expected from the worst case unitaries according to \cite{burgarth_quantum_2022}.
While the numerics in \cite{lee_dependence_2018} do indeed suggest that the bound obtained in this section is not tight, an exact quantum speed limit for the SWAP operation is not yet known.

\subsection{Systems with a single control}
Often in quantum control, one studies control systems with only a single control Hamiltonian
\begin{equation}
    H(t) = H_d + f(t)H_c.
\end{equation}
Almost all such systems are universal \cite{dalessandro_introduction_2021}, i.e. \(\mathfrak{g} = \mathrm{SU}(d)\) or \(\mathfrak{g} = \mathrm{U}(d)\).
The control Hamiltonian \(H_c\) itself becomes a symmetry of \(\mathfrak{h}\).
From \cref{thm:speed_limit_unitary}(b) it now follows that the time to reach any unitary \(U\) is bounded from below by
\begin{equation}
    T \geq \frac{\lVert [U, H_c] \rVert_F \sigma_\mathrm{min}}{2 \lVert H_c \rVert_F \lVert [H_c, H_d] \rVert_F}.
\end{equation}
This demonstrates that in quantum control systems, non-commutativity is indeed an important resource. It does not only affect the dimension of the dynamical Lie algebra, but it also dictates the timescales on which the system can be controlled. Giving it a very handwavy interpretation: the less commutative the target with the control, the more the drift needs to do, leading to a longer time, and the less commutative the control and the drift, the faster the generation of new directions in the dynamics, and therefore the smaller the time.
Interestingly, judging by the bound, the smallest nonzero gap in the spectrum of the control Hamiltonian might play an important role, which has not yet been studied this far.

\subsection{Globally controlled Rydberg atom simulation of the Ising model}
Optically trapped Rydberg atoms have emerged as a popular platform for quantum simulation in recent years (see, e.g. \cite{schaus_observation_2012, bernien_probing_2017, keesling_quantum_2019, surace_lattice_2020, michel_hubbard_2024}).
The van-der-Waals interaction between the atoms plays the role of the drift and is of the form
\begin{equation}
    H_d = \frac{1}{2} \sum_{i \neq j = 1}^N \frac{C}{\lvert \vec{r}_i - \vec{r}_j \rvert ^6} \ket{1}\bra{1}_i \ket{1}\bra{1}_j,
\end{equation}
where \(C\) is a coupling constant and \(\vec{r}_i\) denotes the position of the i'th atom. We want to consider the case here where the Rydberg atoms are trapped in a regular, linear array like in \cite{bernien_probing_2017}, i.e. \(\lvert \vec{r}_i - \vec{r}_j \rvert = a \lvert i - j \rvert\) for some lattice spacing \(a\).
Globally applied laser fields are used to control the system
\begin{equation}
    H_1 = \sum_{i=1}^N X_i \text{, } \quad H_2 = \sum_{i=1}^N Z_i.
\end{equation}

One can verify with the methods from \cite{zimboras_symmetry_2015} for small systems that such a simulator can in fact simulate a 1D Ising model with both transversal and longitudinal fields
\begin{equation}
    H_s = J \sum_{i = 1}^{N-1}Z_i Z_{i+1} + g \sum_{i=1}^N X_i + h \sum_{i=1}^N Z_i, 
\end{equation}
which can be used for example to study dynamical phase transitions \cite{heyl_quenching_2017}.

Since the controls are global, they are left invariant by permutations of the atoms. We can choose for example the SWAP between the first two atoms in the array as the symmetry \(S\).
This symmetry is restored to the drift by modifying the couplings of these two atoms to the rest of the array such that they become indistinguishable to the array.
This perturbation satisfies \(\lVert \Delta H \rVert_\infty = \frac{C}{2a^6} (1 - \frac{1}{(N-1)^6})\) (for a detailed derivation, see \ref{app:rydberg}).

\begin{figure}
    \centering
    \includegraphics[scale=1]{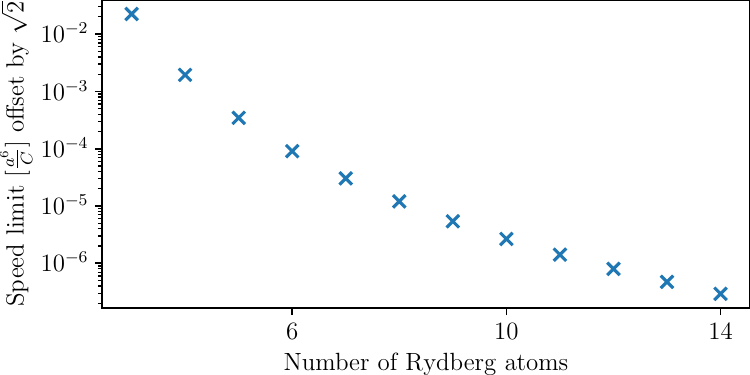}
    \caption{Numerical evaluation of the speed limit bound from \cref{thm:speed_limit_hamiltonian}(b) for the simulation of an Ising model with both transversal and longitudinal fields on a globally controlled Rydberg atom simulator.
    The bound was calculated using the Chebyshev method, see \ref{app:chebyshev}.
    As the number of atoms increases, the speed limit  approaches the value of \(\sqrt{2}\).}
    \label{fig:ising_speed_limit}
\end{figure}

We use the Chebyshev filtering method to numerically evaluate the bound in \cref{thm:speed_limit_hamiltonian}(b) from \(N=3\) up to \(N=14\) for the parameters \(J=1\), \(g=h=\frac{1}{2}\).
The results are depicted in \fref{fig:ising_speed_limit}.
The numerics suggest that the time needed to simulate the Ising model with transverse and longitudinal fields is limited by
\begin{equation}
    T \geq \frac{\sqrt{2}a^6}{C}.
\end{equation}

\subsection{Nuclear magnetic resonance simulation of the Sachdev–Ye–Kitaev model}

The Sachdev-Ye-Kitaev (SYK) model \cite{sachdev_gapless_1993, noauthor_alexei_nodate} gained attention in the condensed matter and high energy physics communities (see e.g. \cite{sachdev_bekenstein-hawking_2015, inkof_quantum_2022, salvati_superconducting_2021}) for its non-Fermi liquid behavior and holographic duality to extremal charged black holes in \(\mathrm{AdS}_2\) geometry.

The Hamiltonian is given by
\begin{equation}
    H = \frac{J_{ijkl}}{4!} \chi_i \chi_j \chi_k \chi_l + \frac{\mu}{4} C_{ij} C_{kl} \chi_i \chi_j \chi_k \chi_l,
\end{equation}
where the \(\chi_i\) are Majorana fermion operators and \(J_{ijkl}\) and \(C_{ij}\) are antisymmetric tensors, which are usually drawn randomly from a Gaussian distribution.

The dynamics of the SYK model has recently been simulated on a molecule via \ac{NMR} \cite{luo_quantum_2019}.
To do this, the Majorana fermion operators are mapped to spin operators via the Jordan Wigner transformation
\begin{eqnarray}
    &\chi_{2i-1} = \frac{1}{\sqrt{2}} X_1X_2 \cdots X_{i-1} Z_i \\
    & \chi_{2i} = \frac{1}{\sqrt{2}} X_i X_2 \cdots X_{i-1} Y_i.
\end{eqnarray}
\ac{GRAPE} can then be used to find \ac{NMR} pulses which drive the nuclear spins of a Crotonic acid molecule in such a way to implement the time evolution under the transformed SYK model.

We use the Chebyshev method (see \ref{app:chebyshev}) to evaluate the bound from \cref{thm:speed_limit_hamiltonian}(b) and optimize the choice of the symmetry \(S\) numerically in order to obtain a lower bound on the necessary pulse duration of \(T \geq \unit[7.6]{ms}\) for 100 random realizations of the SYK model with parameters chosen from a Gaussian of zero mean and unit variance.
This lies well below the \(\unit[100]{ms}\) pulse duration employed in \cite{luo_quantum_2019}, and it would be interesting to see if this is arises from our bound being loose, or from scope for further time optimization of the optimal pulse used.
We observe that the obtained speed limits only differ in the \(\unit{\mu s}\)-regime from each other for different realizations.

%% file: Conclusion.tex
\section{Conclusion}\label{sec:conclusion}
We generalized the distance to uncontrollability approach to quantum speed limits from \cite{burgarth_quantum_2022} to a \textit{distance to unreachability} approach, which factors in the structure of the specific use case of the quantum system.
On the one hand side, this allowed us to give explicit quantum speed limits for specified unitary quantum gates on different control platforms.
On the other hand, we also extended the possible use cases to studying the unitary time evolution of a specified Hamiltonian. Our results facilitate deriving lower bounds on the time needed to implement the worst-case unitary in the orbit of the target Hamiltonian.
Furthermore, we gave examples of systems where the exact quantum speed limit is unknown and compared our bounds to the existing literature.
We encourage experimentalists to evaluate their quantum control systems with our methods to get a sense of where the fundamental limitations and bottlenecks of their hardware platforms lie.

%% file: Appendix.tex
\section{Duhamels principle}
\label{app:duhamel}

Consider two unitary one-parameter groups \(U_1(t)\) and \(U_2(t)\), which are solutions to the differential equations
\begin{eqnarray}
	\frac{\mathrm{d}}{\mathrm{d} t} U_1(t) &= -i H_1(t) U_1(t) \\
	\frac{\mathrm{d}}{\mathrm{d} t} U_2(t) &= -i H_2(t) U_2(t),
\end{eqnarray}
where \(H_1(t)\) and \(H_2(t)\) are Hermitian at any time \(t\).

\begin{lemma}
\label{lem:unitary_bound}
\(\lVert U_1(t) - U_2(t) \rVert_\infty \leq \int_0^t \mathrm{d}s \lVert H_1(s) - H_2(s) \rVert_\infty.\)
\end{lemma}

\begin{proof}
Application of the product rule shows that
\begin{equation}
\frac{\mathrm{d}}{\mathrm{d} t} \left(U_2^\dagger(t) U_1(t)\right) = -i U_2^\dagger(t)  \left[H_1(t) - H_2(t)\right]U_1(t)
\end{equation}
holds true. One can use this to bound the difference between the two one-parameter groups at any time \(t\) by the triangle inequality \cite{nielsen_quantum_2006}
\begin{eqnarray}
	\lVert U_1(t) - U_2(t) \rVert_\infty &= \lVert U_2^\dagger(t)U_1(t) - \mathds{1} \rVert_\infty = \lVert \int_0^t \mathrm{d}s \frac{\mathrm{d}}{\mathrm{d}s} \left(U_2^\dagger(s)U_1(s)\right) \rVert_\infty \nonumber \\
	&= \lVert \int_0^t \mathrm{d}s U_2^\dagger(s)  \left[H_1(s) - H_2(s)\right]U_1(s) \rVert_\infty \nonumber \\
	&\leq \int_0^t \mathrm{d}s \lVert U_2^\dagger(s)  \left[H_1(s) - H_2(s)\right]U_1(s) \rVert_\infty \\
	&= \int_0^t \mathrm{d}s \lVert H_1(s) - H_2(s) \rVert_\infty.
\end{eqnarray}
This statement is also a direct consequence of the Duhamel formula \cite{dellantonio_lectures_2015}.
\end{proof}

For any target unitary \(U\) and control system with reachable set \(\mathcal{R}' \reflectbox{ $\notin$ } U \), we define the following quantity
\begin{equation}
	\delta = \inf_{V \in \mathcal{R}'} \lVert U - V \lVert_\infty.
\end{equation}
Assume that \(H_1(t) = H_d + \sum_{j=1}^n f_j(t) H_j\) and \(H_2(t) = H_d + \Delta H + \sum_{j=1}^n f_j(t) H_j\). Choose control functions \(f_j(t) \text{, } j = 1, ..., n\) for which \(U_1(T) = U\).
Denote with \(U_2(T)\) the unitary obtained by applying these controls on the perturbed control system instead.
Then the chain of inequalities
\begin{equation}
\delta \leq \lVert U_1(T) - U_2(T) \rVert_\infty \leq \int_0^{T} \mathrm{d}s \lVert \Delta H \rVert_\infty \leq \lVert \Delta H \rVert_\infty T
\end{equation}
arises from applying \cref{lem:unitary_bound}.
Therefore we get a lower bound
\begin{equation}
 T \geq \frac{\delta}{\lVert \Delta H \rVert_\infty} \text{ } \forall \Delta H \text{ s.t. } U \notin \mathcal{R}'
\end{equation}
on the time in which \(U\) can be implemented.

\section{Some technical Lemmas}
\label{app:lemmas}

\begin{lemma}
	\label{lem:lower_bound_evolution_from_identity}
	Let \(A\) be a Hermitian matrix, \(v\) be any vector. Then
	\begin{equation}
		\underset{t\in \mathds{R}}{\sup} \lVert \left(e^{-itA} - \mathds{1} \right) v \rVert_2^2 \geq 2 \lVert (\mathds{1} - P_{\mathrm{ker} \, A}) v \rVert_2^2.
	\end{equation}
\end{lemma}

\begin{proof}
	Since \(A\) is Hermitian, it has a orthonormal basis of eigenvectors \(v_k\) with \(A v_k = \lambda_k v_k\).
	Decomposing \(v\) in terms of this basis yields
	\begin{eqnarray}
		\lVert \left(e^{-itA} - \mathds{1}\right) v\rVert_2^2 & = \lVert \sum_k c_k \left(e^{-it\lambda_k} - 1\right) v_k \rVert_2^2 = \sum_k \lvert c_k \rvert^2  \lvert e^{-it\lambda_k} - 1 \rvert^2 \\
		 & = 2 \left(\sum_k \lvert c_k \rvert^2 - \sum_k \lvert c_k \rvert^2 \cos(\lambda_k t)\right), \label{eq:evoultion_from_identity_intermediate}
	\end{eqnarray}
	where the orthonormality of the basis was used for the Pythagorean theorem in the second step.
	Taking the supremum over \(t\) of the expression above is difficult without knowledge of the eigenvalues \(\lambda_k\) of \(A\).
	However, one can use the fact that the infimum of any function is less than or equal to its average value.
	This leads to a lower bound given by the norm of the component of \(v\) outside the kernel of \(A\)
	\begin{eqnarray}
		& \underset{t\in \mathds{R}}{\sup} \lVert \left(e^{-itA} - \mathds{1} \right) v \rVert_2^2 = \underset{T > 0}{\sup} \underset{t\in \left[-T, T\right]}{\sup} \lVert \left(e^{-itA} - \mathds{1} \right) v \rVert_2^2 \\
		\overset{\eref{eq:evoultion_from_identity_intermediate}}{=} & \underset{T > 0}{\sup} \text{ } 2 \left(\sum_k \lvert c_k \rvert^2 - \underset{t \in \left[-T, T\right]}{\inf} \sum_k \lvert c_k \rvert^2 \cos(\lambda_k t)\right) \\
	 \geq &  \underset{T > 0}{\sup} \text{ } 2 \left(\sum_k \lvert c_k \rvert^2 -  \sum_k \lvert c_k \rvert^2 \frac{1}{2T} \int_{-T}^T \mathrm{d}t \cos(\lambda_k t)\right) \\
		\geq & \underset{T > 0}{\sup} \text{ } 2 \left(\sum_{k: \lambda_k \neq 0} \lvert c_k \rvert^2 -  \sum_{k: \lambda_k \neq 0} \lvert c_k \rvert^2 \frac{1}{\lvert \lambda_k \rvert T}\right) = 2 \sum_{k: \lambda_k \neq 0} \lvert c_k \rvert^2 \label{eq:evolution_from_identity_supremum_lower_bound}.
	\end{eqnarray}
\end{proof}

\begin{lemma}
	\label{lem:lower_bound_projection_to_kernel}
	Let \(A\) be a \(d\times d\) matrix, \(v\) be any \(d\)-dimensional vector. Then
	\begin{equation}
		\lVert (\mathds{1} - P_{\mathrm{ker} \, A}) v \rVert_2^2 \geq \max\left\{\frac{\langle v, Av \rangle}{\lVert A \rVert_\infty}, \frac{\lVert Av \rVert_2^2}{\lVert A \rVert_\infty^2}\right\}.
	\end{equation}
\end{lemma}

\begin{proof}
    We want to provide lower bounds to the component of \(v\), which lies outside the kernel of \(A\), without having to explicitly compute said kernel.
	One way to do this is via
	\begin{equation}
		\langle v, Av \rangle = \sum_k \lvert c_k \rvert^2 \lambda_k \leq \lVert A \rVert_\infty \sum_{k: \lambda_k \neq 0} \lvert c_k \rvert^2,
	\end{equation}
	or alternatively by
	\begin{equation}
		\lVert Av \rVert_2^2 = \sum_k \lvert \lambda_k \rvert^2 \lvert c_k \rvert^2 \leq \lVert A \rVert_\infty^2 \sum_{k: \lambda_k \neq 0} \lvert c_k \rvert^2.
	\end{equation}
\end{proof}

\begin{lemma}
    \label{lem:adjoint_hermitian}
	For any (finite-dimensional) Hilbert space \(\mathcal{H}\) consider the Hilbert space of linear operators on \(\mathcal{H}\) equipped with the Hilbert-Schmidt scalar product \(B_{\mathrm{HS}}(\mathcal{H})\).
	Then the adjoint representation of \(H\) on \(B_{\mathrm{HS}}(\mathcal{H})\) as \(\text{ad}_{H}(A) = [H, A] \text{, } A \in B_{\mathrm{HS}}(\mathcal{H}) \) is a Hermitian operator on \(B_{\mathrm{HS}}(\mathcal{H})\).
\end{lemma}

\begin{proof}
	For \(A, B \in B_{\mathrm{HS}}(\mathcal{H})\) consider
	\begin{eqnarray}
		\langle B, \text{ad}_H(A) \rangle &= \text{Tr}(B^\dagger [H, A]) = \text{Tr}(B^\dagger H A - B^\dagger A H) \nonumber \\
        &= \text{Tr}(B^\dagger H A - H B^\dagger A) = \text{Tr}([B^\dagger , H]A) = \text{Tr}([H, B]^\dagger A) \nonumber \\
		& = \langle \text{ad}_H(B), A \rangle,
	\end{eqnarray}
	where linearity and cyclicity of the trace were used in the third equality. 
\end{proof}

\begin{lemma}
\label{lem:rotated_unitary_bound}
	Let \(U\) be a unitary matrix such that \(\lVert U - \mathds{1} \rVert < \epsilon\) in any unitarily invariant norm and X be an arbitrary Hamiltonian.
	Then
	\begin{equation}
		\lVert e^{-iUXU^\dagger t} - e^{-iXt} \rVert < 2 \epsilon \text{ } \forall t \in \mathds{R}.
	\end{equation}
\end{lemma}
\begin{proof}
	\begin{equation}
        \eqalign{
    		\fl \left\lVert e^{-iUXU^\dagger t} - e^{-iXt} \right\rVert = \left\lVert Ue^{-iXt}U^\dagger + Ue^{-iXt} - Ue^{-iXt} - e^{-iXt} \right\rVert \\
    		\leq \left\lVert Ue^{-iXt}U^\dagger - Ue^{-iXt} \right\rVert + \left\lVert Ue^{-iXt} - e^{-iXt} \right\rVert = 2 \lVert U - \mathds{1} \rVert < 2 \epsilon,
        }
	\end{equation}
	where the second equality is due to the unitary invariance of the norm.
\end{proof}

\section{Chebyshev polynomial spectral filtering}
\label{app:chebyshev}

The goal is to find a lower bound on \(\lVert (\mathds{1} - P_{\mathrm{ker}\, A}) v \rVert_2\) for an Hermitian matrix \(A\) and some vector \(v\) for which we do not need to fully diagonalize A.
Without loss of generality we are going to assume that \(A\) is positive semidefinite, as for any Hermitian matrix \(A\) we have that \(P_{\mathrm{ker}\, A} = P_{\mathrm{ker}\, A^2}\) and \(A^2\) is positive semidefinite.
Denote with \(\sigma_\mathrm{min}\) the smallest nonzero eigenvalue of \(A\) and \(\sigma_\mathrm{max}\) the largest eigenvalue of \(A\).
Let \(p\) be a polynomial such that \(p(0) = 1\) and \(\lvert p(x) \rvert \leq \epsilon\) for \(x \in [\sigma_\mathrm{min}, \sigma_\mathrm{max}]\). Then, we have that \(\lVert P_{\mathrm{ker}\, A} - p(A) \rVert_\infty \leq \epsilon\). Therefore we can make use of the reverse triangle inequality to get that
\begin{eqnarray}
    \lVert (\mathds{1} - P_{\mathrm{ker}\, A}) v \rVert_2 & \geq \lVert v \rVert_2 - \lVert P_{\mathrm{ker}\, A} v \rVert_2  \nonumber \\
    & = \lVert v \rVert_2 - \lVert \left(P_{\mathrm{ker}\, A} - p(A) + p(A)\right)  v \rVert_2 \\
    & \geq (1-\epsilon) \lVert v \rVert_2 - \lVert p(A)v \rVert_2. \nonumber
\end{eqnarray}
Now it only remains to construct a suitable polynomial. It is well-known that shifted Chebyshev polynomials $T_m$ are the optimal choice in the sense that they minimize the error \(\epsilon\) among all polynomials with a fixed degree \cite{flanders_numerical_1950, golub_chebyshev_1961, gutknecht_chebyshev_2002}.
Choosing
\begin{equation}
    p(x) = \frac{T_m\left(\frac{\sigma_\mathrm{max} + \sigma_\mathrm{min} - 2x}{\sigma_\mathrm{max} - \sigma_\mathrm{min}}\right)}{T_m \left(\frac{\sigma_\mathrm{max} + \sigma_\mathrm{min}}{\sigma_\mathrm{max} - \sigma_\mathrm{min}}\right)}
\end{equation}
yields an approximation error
\begin{equation}
    \epsilon \leq \mathrm{sech}\left(m \cosh\left(\frac{\sigma_\mathrm{max} + \sigma_\mathrm{min}}{\sigma_\mathrm{max} - \sigma_\mathrm{min}}\right)\right),
\end{equation}
which vanishes exponentially in \(m\).

We want to remark that it is not strictly necessary to know the exact values of \(\sigma_\mathrm{min}\) and \(\sigma_\mathrm{max}\), as all the statements above remain true if one inserts upper bounds to \(\sigma_\mathrm{max}\) and lower bounds to \(\sigma_\mathrm{min}\) instead.
A trivial bound to \(\sigma_\mathrm{max}\) is for example given by \(\sigma_\mathrm{max} \le \lVert A \rVert_F\).
Finding lower bounds to \(\sigma_\mathrm{min}\) is however still difficult in general.

\section{Details on the SWAP example}
\label{app:swap}

A suitable symmetry operator will be constructed through methods similar to those detailed in \cite{burgarth_quantum_2022}.
The crucial observation is that the control only acts on a one-dimensional subspace.
Therefore the Hilbert space can be split into \(\mathcal{H} = \mathcal{H}_c \oplus \mathcal{H}_c^\perp\), where \(\mathcal{H}_c = \mathds{C} \ket{1}\bra{1}\) is left invariant by the control.
Now assume that the perturbation \(\Delta H\) introduces a degeneracy into \(H_d\).
This means that one can construct an eigenstate of \(H_d\) which has no support on \(\mathcal{H}_c\).
Therefore it is left invariant by the control and the projector onto that eigenstate will be a symmetry of the perturbed control system.

To make this explicit, first use the fact that \(H_d\) is a tridiagonal Toeplitz matrix, which enables one to directly write down its spectrum
\begin{equation}
	E_k = 2J\cos\left(\frac{\pi k}{N+1}\right)\text{, } k=1,...,N
\end{equation}
and eigenstates
\begin{equation}
	\ket{\alpha_k} = \sqrt{\frac{2}{N+1}} \sum_{m=1}^{N} \sin\left(\frac{\pi k m}{N+1}\right)\ket{m}\text{, } k=1,...,N \label{eq:n_level_eigenstates}
\end{equation}
according to \cite{noschese_tridiagonal_2013}.
The closest energy gap is between \(E_1\) and \(E_2\), which can be upper bounded by \(\lvert E_1 - E_2 \rvert < \frac{3 \pi^2}{N^2}J\) \cite{burgarth_quantum_2022}.
Therefore, the operator norm of the perturbation \(\Delta H = (E_1 - E_2) \ket{\alpha_2}\bra{\alpha_2}\) is bounded from above by the same bound.

The corresponding, now degenerate, eigenstates
\begin{eqnarray}
	& \ket{\alpha_1} = \sqrt{\frac{2}{N+1}}\sum_{m=1}^{N} \sin\left(\frac{\pi m}{N+1}\right)\ket{m} \\
	& \ket{\alpha_2} = \sqrt{\frac{2}{N+1}}\sum_{m=1}^{N} \sin\left(\frac{\pi 2 m}{N+1}\right)\ket{m}
\end{eqnarray}
need to be combined in such a way that they no longer overlap the \(\ket{1}\) state.
This is achieved by
\begin{equation}
	\ket{\alpha} = \frac{1}{\sqrt{1 + \frac{\sin^2\left(\frac{2 \pi}{N+1}\right)}{\sin^2\left(\frac{\pi}{N+1}\right)}}}\left(\ket{\alpha_2} - \frac{\sin\left(\frac{2 \pi}{N+1}\right)}{\sin\left(\frac{\pi}{N+1}\right)} \ket{\alpha_1}\right), \label{eq:n_level_symmetry_projector}
\end{equation}
which does satisfy \(\braket{1}{\alpha} = 0\).
Therefore, \(S = \ket{\alpha}\bra{\alpha}\) will be the symmetry used to derive the speed limit.

The SWAP operator does indeed break this symmetry, as
\begin{equation}
    \label{eq:swap_symmetry_commutator}
    \eqalign{
    	C & \defeq [\text{SWAP}, S] \\
        & = \ket{1}\braket{N}{\alpha} \bra{\alpha} - \ket{N} \braket{N}{\alpha} \bra{\alpha} - \ket{\alpha}\braket{\alpha}{N}\bra{1} + \ket{\alpha} \braket{\alpha}{N} \bra{N} \\
    	& = \braket{N}{\alpha} \left(\ket{1}\bra{\alpha} - \ket{N}\bra{\alpha}\right) - \braket{\alpha}{N} \left(\ket{\alpha}\bra{1} - \ket{\alpha} \bra{N}\right) \neq 0.
    }
\end{equation}

In order to calculate the Frobenius norm of the commutator, use that \(C^\dagger = -C\), which can be observed from \eref{eq:swap_symmetry_commutator}.
Therefore
\begin{equation}
    \fl C^\dagger C = -C^2 = \lvert \braket{N}{\alpha} \lvert^2 \left(\ket{1}\bra{1} + \ket{N}\bra{N} + 2 \ket{\alpha}\bra{\alpha}\right) \text{+ offdiagonal terms}. \label{eq:swap_symmetry_commutator_square}
\end{equation}
Since the trace is independent of the chosen basis and \(\{\ket{1}, \ket{\alpha}, \ket{N}\}\) are linearly independent, they can be completed to a basis and the trace can be calculated with respect to that basis.
Thus,
\begin{equation}
	\lVert [\text{SWAP}, S] \rVert_F = \sqrt{\text{Tr}\left(C^\dagger C\right)} \overset{\eref{eq:swap_symmetry_commutator_square}}{=} 2 \lvert \braket{N}{\alpha} \rvert.
\end{equation}
By the same argument it is also easily seen that \(\lVert S \rVert_F = 1\).

The only thing remaining to calculate is \(\lvert \braket{N}{\alpha}\rvert\), which is straightforward to do using \eref{eq:n_level_eigenstates} and \eref{eq:n_level_symmetry_projector}
\begin{eqnarray}
	\fl \lvert \braket{N}{\alpha} \rvert = \frac{1}{\sqrt{1 + \frac{\sin^2\left(\frac{2 \pi}{N+1}\right)}{\sin^2\left(\frac{\pi}{N+1}\right)}}} \sqrt{\frac{2}{N+1}} \left\lvert \left( \sin\left(\frac{2 \pi N}{N+1}\right)  - \frac{\sin\left(\frac{2 \pi}{N+1}\right)}{\sin\left(\frac{\pi}{N+1}\right)} \sin\left(\frac{\pi N}{N+1}\right) \right)\right\rvert.
\end{eqnarray}

Collecting all results and inserting them into \cref{thm:speed_limit_unitary}(b) finally yields an expression for the speed limit of the SWAP operation
\begin{eqnarray}
	\fl T_\text{SWAP} \geq \frac{N^2}{3\pi^2 J \sqrt{1 + \frac{\sin^2\left(\frac{2 \pi}{N+1}\right)}{\sin^2\left(\frac{\pi}{N+1}\right)}}} \sqrt{\frac{2}{N+1}} \left\lvert \sin\left(\frac{2 \pi N}{N+1}\right) - \frac{\sin\left(\frac{2 \pi}{N+1}\right)}{\sin\left(\frac{\pi}{N+1}\right)} \sin\left(\frac{\pi N}{N+1}\right)\right\rvert \nonumber \\
	= \frac{2 N^2}{3\pi^2 J \sqrt{3 + 2\cos\left(\frac{2 \pi}{N + 1}\right)}} \sqrt{\frac{2}{N+1}} \left \lvert \sin\left(\frac{2 \pi N}{N+1}\right) \right\rvert.
\end{eqnarray}

\section{Details on the Rydberg simulator example}
\label{app:rydberg}

Here, we find the operator norm of a perturbation \(\Delta H\) that restores the SWAP symmetry between the first two atoms to the van-der-Waals interaction in the linear array of Rydberg atoms.
One can for example replace the couplings of atom one and two to any atom \(j\) in the remaining array with the average between the two.
This perturbation is given explicitly as
\begin{equation}
    \eqalign{
        \Delta H &= \sum_{j=3}^N \frac{C}{a^6}\left(\frac{1}{2} \left(\frac{1}{\lvert j - 1 \rvert^6} + \frac{1}{\lvert j - 2 \rvert^6}\right) - \frac{1}{\lvert j - 1 \rvert^6}\right) \ket{1}\bra{1}_1 \ket{1}\bra{1}_j\\
        & \hphantom{=} + \sum_{j=3}^N \frac{C}{a^6}\left(\frac{1}{2} \left(\frac{1}{\lvert j - 1 \rvert^6} + \frac{1}{\lvert j - 2 \rvert^6}\right) - \frac{1}{\lvert j - 2 \rvert^6}\right) \ket{1}\bra{1}_2 \ket{1}\bra{1}_j \\
        & = \frac{C}{2a^6} \ket{1}\bra{1}_1 \sum_{j=3}^N \left(\frac{1}{\lvert j-2\rvert^6} - \frac{1}{\lvert j- 1 \rvert^6}\right) \ket{1}\bra{1}_j \\
        & \hphantom{=} + \frac{C}{2a^6} \ket{1}\bra{1}_2 \sum_{j=3}^N \left(\frac{1}{\lvert j-1\rvert^6} - \frac{1}{\lvert j - 2 \rvert^6}\right) \ket{1}\bra{1}_j.
    }
\end{equation}

The perturbation is diagonal in the \(Z\) basis. The eigenstates that belong to the eigenvalues with the largest modulus are \(\ket{101\cdots1}\) and \(\ket{011\cdots1}\).
Hence, we get for the operator norm
\begin{eqnarray}
    \lVert \Delta H \rVert_\infty & = \frac{C}{2a^6} \sum_{j=3}^N \left(\frac{1}{\lvert j - 2 \rvert^6} - \frac{1}{\lvert j - 1 \rvert^6}\right) = \frac{C}{2a^6} \sum_{j=3}^N \frac{((j-1)^6-(j-2)^6)}{(j-1)^6(j-2)^6} \nonumber \\
    & = \frac{C}{2a^6} - \frac{C}{2a^6(N-1)^6}.
\end{eqnarray}